\documentclass[onecolumn]{emulateapj}
\usepackage{amsmath}
\usepackage{amssymb}
\usepackage{amstext}
\usepackage{graphicx}

\newcommand{\be}{\begin{displaymath}}
\newcommand{\ee}{\end{displaymath}}
\newcommand{\bea}{\begin{eqnarray}}
\newcommand{\eea}{\end{eqnarray}}

\begin{document}

\shorttitle{ACCRETION ONTO FAST X-RAY PULSARS}
\shortauthors{RAPPAPORT, FREGEAU,  \& SPRUIT}
\submitted{Submitted to ApJ}
\title{Accretion Onto Fast X-Ray Pulsars}

\author{S.\ A.\ Rappaport\altaffilmark{1}, J.\ M.\ Fregeau\altaffilmark{1},  \& H. \ Spruit\altaffilmark{2}}

\altaffiltext{1}{Department of Physics, MIT 37-602b, 77 Massachusetts Ave, 
                 Cambridge, MA 02139; {\tt sar@mit.edu}}
\altaffiltext{2}{Max-Planck-Institut f\"ur Astrophysik, Box 1317, 85741, Garching, Germany; {\tt henk@mpa-Garching.mpg.de}}

\setlength{\parskip}{8pt}

\begin{abstract}

The recent emergence of a new class of accretion-powered, transient, millisecond X-ray pulsars presents some difficulties for the conventional picture of accretion onto rapidly rotating magnetized neutron stars and their spin behavior during outbursts.   In particular, it is unclear from the standard paradigm how these systems manage to accrete over such a wide range in $\dot M$ (i.e., $\gtrsim$ a factor of 50), and why the neutron stars exhibit a high rate of {\em spindown} in at least a number of cases.   Following up on prior suggestions, we propose that `fast' X-ray pulsars can continue to accrete, and that their accretion disks terminate at approximately the corotation radius.  We demonstrate the existence of such disk solutions by modifying the Shakura-Sunyaev equations with a simple magnetic torque prescription.  The solutions are completely analytic, and have the same dependence on $\dot M$ and $\alpha$ (the viscosity parameter) as the original Shakura-Sunyaev solutions; but, the radial profiles can be considerably modified, depending on the degree of fastness.  We apply these results to compute the torques expected during the outbursts of the transient millisecond pulsars, and find that we can explain the large spindown rates that are observed for quite plausible surface magnetic fields of $\sim 10^9$ G. 

\end{abstract}

\keywords{accretion disks --- accretion torques --- binaries --- magnetic fields --- pulsars --- stars: neutron --- X-rays: binaries}

\section{Introduction}

The recent discovery of five accretion-powered millisecond X-ray pulsars (SAX J1808-3658; XTE J0929-314; XTE J1751-305; XTE 1807-294; XTE J1814-338; Wijnands \& van der Klis 1998; Chakrabarty \& Morgan 1998; Galloway et al. 2002; Markwardt et al. 2002; Markwardt, Smith, \& Swank 2003; Markwardt, Juda, \& Swank 2003; Markwardt, \& Swank 2003; Markwardt, Strohmayer, \& Swank 2003), all of which are X-ray transients, raises a number of interesting questions about the fundamentals of accretion onto rapidly rotating neutron stars.  These 5 msec pulsars have typical luminosities that range from about $\sim 5 \times 10^{36}$ ergs s$^{-1}$ near the start of the X-ray outbursts, down to as low as $\sim 5 \times 10^{34}$ ergs s$^{-1}$ (for assumed distances of $\sim$5 kpc) as they fade in intensity below the detection limit (e.g., of the RXTE satellite).   The X-ray pulsations are typically detected as long as any DC intensity is detected from the source.  The pulse profiles are basically sinusoidal and do not change dramatically in character as the source intensity drops.  In particular, there is no evidence of anything like a ``propeller effect'' setting in as the intensity gradually falls to quite low values.  

These observational facts raise a number of puzzling issues (see also, e.g., Burderi \& King 1998; Psaltis \& Chakrabarty 1999).  First, how are these rapidly rotating, magnetized neutron stars able to accrete matter over such a wide range (e.g., $\sim 50$) in mass accretion rate, $\dot M$, without encountering a ``centrifugal barrier'' effect (e.g., Illarionov \& Sunyaev 1975, hereafter, IS; see also Davidson \& Ostriker 1973; Fabian 1975; Lipunov \& Shakura 1976)?  Second, why do at least two of the msec X-ray pulsars exhibit a large {\em spindown} in pulse period, i.e., $\dot P > 0$, as the luminosity declines throughout the outburst (Galloway et al. 2002; Morgan, Galloway, \& Chakrabarty 2003)?  Third, why is it that---thus far---only relatively faint X-ray transients have been found to exhibit coherent periodic msec pulsations, while many other ``steady'' low-mass X-ray binaries (LMXBs) do not?  

In the conventional picture of accretion onto a magnetized neutron star via an accretion disk, the magnetospheric radius is taken to be
\bea
r_m \simeq \left( GM\right)^{-1/7}\dot M^{-2/7} \mu^{4/7} 
= 35~ {\rm km} \left(\frac{M}{1.4~M_\odot}\right)^{-1/7} 
\left(\frac{\dot M}{10^{17} {\rm g~s}^{-1}}\right)^{-2/7} \mu_{26.5}^{4/7} ~~~ , \label{rm}
\eea
where $M$ is the mass of the neutron star, $\dot M$ is the steady-state mass accretion rate,  and $\mu$ is the magnetic dipole moment of the neutron star (e.g., Lamb, Pethick, \& Pines 1973 ``LPP''; Rappaport \& Joss 1976; Ghosh \& Lamb 1979 ``GL''; Wang 1987).  The quantity $\mu_{26.5}$ is the dipole moment expressed in units of $10^{26.5}$ G cm$^3$, which corresponds to a surface magnetic field at the poles of $\sim 3 \times 10^8$ G.  Here and throughout this work we take eq. (1) to be a formal definition of $r_m$.  Inside the magnetospheric radius the dynamics of the accreting matter are supposedly dominated by the magnetic field of the neutron star.  The corotation radius is defined as the radial distance at which the Keplerian angular frequency is equal to the spin frequency of the neutron star:
\bea
r_c = \left(\frac{GM}{w_s^2}\right)^{1/3} = 31~{\rm km}\left(\frac{P}{0.003~{\rm s}}\right)^{2/3}
\left(\frac{M}{1.4~M_\odot}\right)^{1/3} ~~~~.
\eea

According to conventional wisdom, the magnetospheric radius, $r_m$, is supposed to lie within the corotation radius, $r_c$ (LPP; GL), in order for accretion to take place.  Otherwise, it is thought that the centrifugal barrier, experienced by particles forced to corotate faster than the Keplerian velocity, would expel matter from the system (IS).   Such pulsars are termed ``fast'', and arise, for a given system, when the accretion rate drops below the value implied by setting $r_c = r_m$.   At the other extreme, if the accretion rate becomes too large, then the value of $r_m$ could drop below a value of $\sim$ 10 km, the approximate radius of the accreting neutron star.  If the accretion disk persists down to the surface of the neutron star, detectable pulsations might not arise.  Thus, for typical system parameter values associated with the msec X-ray pulsars, $r_m$ might have to lie within a rather restricted range, e.g., $10-35$ km.  A range of $3-4$ in $r_m$ corresponds to a range in $\dot M$ of about a factor of 100 (see eq. [1]).   However, the observed range in X-ray luminosity, and presumably also in $\dot M$, is of order a factor of $\sim 50$ for at least three of the msec pulsars.  Thus, in the standard accretion paradigm the magnetic field and values of $\dot M$ would have to be in just the correct range to allow accretion to continue throughout the transient outburst (see also Burderi \& King 1998; Psaltis \& Chakrabarty 1999).  Of course, this is still plausible as the result of observational selection effects, i.e., the sources we observe to pulse have just the right parameters; however, the range of allowable parameter space is growing uncomfortably small.  Moreover, the neutron star should be observed to be spinning up during most of the outburst, while the opposite was true on at least one occasion each, for two of the sources (SAX J1808-3658 and XTE J0929-314).  

The conventional ``propeller'' picture, in which all mass is expelled
as soon as the nominal magnetosphere radius lies outside the
corotation radius (IS), is known to be incorrect from a
theoretical point of view. Spruit and Taam (1993, hereafter ST)
have pointed out that the velocity excess of the stellar rotation
over the Keplerian velocity at $r_{\rm m}$ is energetically
insufficient to gravitationally unbind all accreting matter, unless
$r_{\rm m}/r_{\rm c}$ is actually rather large. Hence there
must be a regime with  $r_{\rm m}\gtrsim r_{\rm c}$ where
accretion in the standard way  is not possible, but at the same
time the bulk of the accreting mass cannot leave the system
either. ST showed, with a time-dependent calculation, that what
happens instead is that accretion continues at some level, while
outside $r_{\rm c}$ mass builds up to high surface densities
(more accurately: surface density normalized to accretion rate). 
The high densities in the inner disk cause the {\it net} angular momentum
flux to be directed {\it outward}, while mass accretion continues inward.

In this paper we argue that the accretion disk structure around a fast pulsar will adjust itself so that the inner edge of the disk will remain fixed near $\sim r_c$, and accretion will continue.  Matter will then {\em not} be ejected from the system in substantial quantity by the propeller effect (IS; if the magnetic moment of the neutron star is not perfectly aligned with its rotation axis), in conjunction with a ``centrifugal barrier".   We compute approximate disk solutions that allow for penetration of the disk to $r_c$.  The solutions for the pressure, density, temperature, and disk thickness are in the form of analytic functions of radial distance from the neutron star and also depend on $\dot M$, the viscosity parameter, and the magnetic moment of the neutron star.  

In \S 2 we set up the steady-state equations for a thin accretion disk around a fast X-ray pulsar and find analytic solutions.  Torques on the neutron star due to accretion and magnetic drag on the disk are evaluated in \S 3.  We summarize our results and draw conclusions in \S 4. 

\section{Magnetically Torqued Accretion}

\subsection{Magnetospheric Accretion in the Thin Disk Model}

We now examine whether there exist solutions for a thin disk surrounding a ``fast'' pulsar, where the inner edge of the disk penetrates to the corotation radius $r_c$.  We start with the equation for conservation of angular momentum:
\bea
\frac{1}{2 \pi Hr} \frac{d}{dr} \left(\dot M \Omega r^2 \right) = \tau_\alpha + \tau_B ~~~~ ,
\eea
where $\Omega$ is the local Keplerian frequency in the disk, $H$ is the full disk thickness, $\tau_\alpha$ is the viscous torque per unit volume, and $\tau_B$ is the magnetic torque on the disk per unit volume.   In words, this equation says that the angular momentum transported inward by matter flow is driven by the viscous and magnetic torques, where a positive sign for $\tau$ here indicates a tendency to slow down the disk rotation.  In the standard ``$\alpha$-viscosity'' prescription, the quantity $\tau_\alpha$ is given by:
\bea
\tau_\alpha = \frac{\alpha}{Hr} \frac{d}{dr} \left( Pr^2H \right) ~~~~ ,
\eea
where $P$ is the pressure in the disk midplane; both $P$ and $H$ are functions of $r$, the radial distance from the neutron star.  Here, $\alpha P$ is the standard viscous stress introduced by Shakura \& Sunyaev (1974; hereafter SS).

Before entering into the details of the model used below for the magnetic interaction between a disk and a magnetosphere, consider first a simpler model (see also ST) which illustrates some properties of the problem that are independent of the detailed model used for the magnetic torques. 

In a standard thin disk accreting onto a {\it slowly} rotating object, the angular momentum flux, $\dot J$, and the mass flux, $\dot M$, are connected directly by the relation $\dot J=\dot Mr_{\rm i}^2\Omega_{\rm i}$, where $r_{\rm i}$ is the radius of the inner edge of the disk and $\Omega_{\rm i}$ the Keplerian rotation frequency at this radius. This is independent of the detailed conditions at the inner edge of the disk, and is a consequence of the fact that the rotation curve has a maximum, near $r_{\rm i}$, where the viscous contrbution to the angular momentum flux vanishes (Lynden-Bell \& Pringle 1972, SS). For accretion onto a rapidly spinning magnetosphere, there is no such extremum, and the angular momentum flux becomes an additional essential parameter of the problem. Its value is determined by the accretion rate and the details of the interaction between disk and magnetosphere, as illustrated in the following simplified problem. 

\subsection{Accretion Torque Applied at Disk Inner Edge}

Consider the case of a rapidly rotating, magnetized neutron star that exerts a torque only over an infinitesimal radial range at the inner edge of the accretion disk -- here taken to be located at $r_c$.  If we approximate the magnetic torque per unit volume as:
\bea
\tau_B =  \frac{\Gamma_c \delta(r-r_c)}{2 \pi r H}
\eea
where $\delta(r-r_c)$ is a Dirac delta function and $\Gamma_c$ is the magnitude of the torque applied by the neutron star at the inner edge of the disk, we find that eq. (3) becomes:
\bea
\frac{1}{2\pi Hr} \frac{d}{dr} \left( \dot M \Omega r^2 \right) = \frac{\alpha}{Hr} \frac{d}{dr} \left( PHr^2 \right) 
+ \frac{\Gamma_c \delta(r-r_c)}{2 \pi r H} ~~~~ ,
\eea
where a {\em negative} value of $\Gamma_c$ represents the torque from a ``fast'' pulsar.  This equation can be directly integrated to yield:
\bea
\dot M \left(\Omega r^2-\Omega_c r_c^2 \right) = 2 \pi \alpha P H r^2 + \Gamma_c  ~~~~ ,
\eea
where the evaluation of the $PHr^2$ term at $r_c$ is set equal to zero under the assumption that the density vanishes inside this point and therefore there can be no viscous torque acting on the innermost edge of the disk.  We identify the combined terms $(\Gamma_c + \dot M \Omega_cr_c^2)$ as the net angular momentum flow into the neutron star, and define their sum to be $j \dot M \Omega_c r_{c}^{2}$, where $j$ is a dimensionless factor.  The solution for the quantity $PH$ can then be written as:
\bea
PH = \frac{\dot M \Omega}{2 \pi \alpha} \left \{1-j\sqrt{\frac{r_c}{r}}\right \} ~~~~.
\eea
Since we are considering only the case of fast pulsars, the factor $j$ can, in principle, take on any value $\leqslant 1$ (i.e., $\Gamma_c \leqslant 0$), the upper limit on $j$ corresponding to zero magnetic torque on the inner edge of the disk.  The physical meaning of this equation can be made clear without solving for the radial profiles of the individual parameters such as density, pressure, disk thickness, and so forth.  This is facilitated by switching explicitly to a viscosity parameter rather than using the standard $\alpha$ prescription (SS).  The relation between these two is: $\alpha P =  - \nu \rho r d\Omega/dr$, where $\nu$ is the coefficient of kinematic viscosity.  With this substitution, and the explicit assumption of Keplerian rotation, eq. (8) becomes:
\bea
\Sigma = \frac{\dot M}{3 \pi \nu} \left \{1-j\sqrt{\frac{r_c}{r}}\right \} ~~~~,
\eea
where $\Sigma$ is the surface column density in the disk.  Note, that in the transformation between $\nu$ and $\alpha$, they cannot both be taken to be constants independent of radial location, so that the radial profile of $\Sigma$ would be different in the two formulations; nonetheless, the qualitative features that we wish to demonstrate will still be present in both.

\begin{figure}
\hskip2.5cm \includegraphics[height=5.0in,angle=-90]{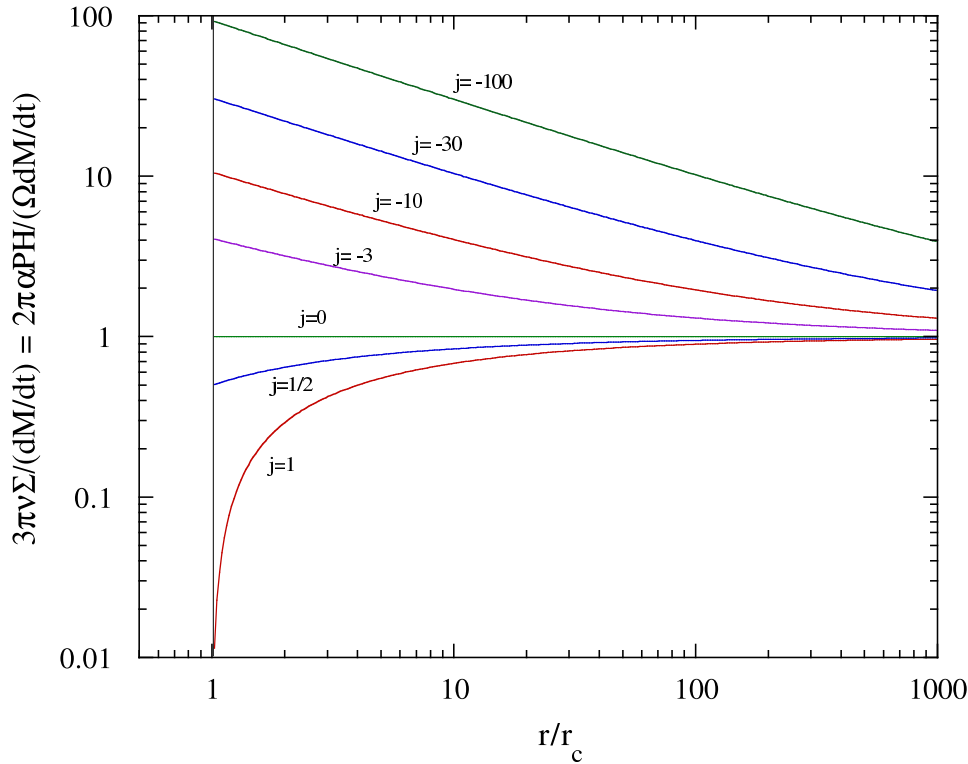}
\caption{\label{fig:thin2}
Normalized surface density distribution $\Sigma(r)$ in a standard thin disk with external torques acting only at the inner edge, i.e., at $r=r_c$. Parameter $j$ is the net specific angular momentum flux through the disk (advection plus magnetic torque) in units of the angular momentum at the inner edge. For accretion onto a slowly rotating object, $j=+1$ applies independent of conditions at the magnetoshere. For accretion onto a rapidly rotating magnetosphere, the angular momentum flux can be outward ($j<0$, spindown of the star), and depends on details of the interaction between magnetosphere and disk.} 
\end{figure}

Figure \ref{fig:thin2} shows $\Sigma(r)$, as a function of the parameter $j$. Note how the surface density near $r_c$ increases dramatically when the angular momentum flux deviates from its standard value of $j=+1$, and then drops abruptly to zero for $r<r_c$. The inner regions of a disk with a net outward angular momentum flux (corresponding to $j<0$) contain much more mass than in the standard case. For the net angular momentum flux (viscous plus advection) to be outward, the viscous flux has to be sufficiently large, which requires a high surface density (for further discussion see ST). 

These lowest-order considerations do not tell us, however, what determines the value of the angular momentum flux -- except in the slow rotator case. A model for the transition region between disk and magnetosphere is needed, which specifies the magnetic torques on the disk in a physically plausible way. Such a model is developed below. One can already expect, however, that the equivalent of Fig. \ref{fig:thin2} will look similar, but with the jump at $r/r_c=1$ broadened, while the parameter $j$ will become a calculable function of the accretion rate.

\subsection{Model for a Disk with Distributed Magnetic Torques}

Here we take the magnetic torque on the disk to be distributed over a range of radial distances.  Spefically, we take the magnetic torque per unit volume to be:
\bea
\tau_B = \frac{B_z B_\phi r}{2 \pi H}  ~~~ .
\eea
Without a full, self-consistent magnetohydrodynamic solution of the disk equations, there is no simple way to determine $B_\phi$.  However, to a level of sophistication consistent with the SS-disk solutions, we simply adopt the following {\em ad hoc} but plausible prescription for $B_\phi$:
\bea
B_\phi \simeq - B_z \left( 1 - \frac{\Omega}{\omega_s} \right) ~~~~~ ,
\eea
which we take to hold for $r > r_c$, and $\omega_s$ is the rotation frequency of the neutron star (see also, Wang 1987; 1995; 1996 for more sophisticated and physically motivated versions of this relation).  Here and in the following we assume, without loss of generality, that $B_z >0$.  This expression has the property that the magnetic torque vanishes at the corotation radius, while $B_\phi$ is comparable to $B_z$ at large distance. The latter is compatible with the fact that $B_\phi$ can not be larger than $B_z$ over significant distances, for reasons of equilibrium and stability of the field above the disk plane.  As the field is wound up by differential rotation  between the star and the disk, the azimuthal component first increases, but the increasing energy in the azimuthal field component pushes the field configuration outward into an open configuration when the azimuthal component becomes comparable to the poloidal component. This was proven by Aly (1984, 1985) in a rather general context, and worked out in some detail for the case of a disk around a magnetic star by Lynden-Bell \& Boily(1994).  Finally, equation (3) reduces to:
\bea
\frac{1}{2\pi Hr} \frac{d}{dr} \left( \dot M \Omega r^2 \right) = \frac{\alpha}{Hr} \frac{d}{dr} \left( PHr^2 \right) 
- \frac{B_z^2 r}{2\pi H} \left( 1-\Omega/\omega_s \right) ~~~~ .
\eea

If we now integrate both sides of equation (12) with respect to $r$, starting at the corotation radius, we find:
\bea
\dot M \left(\Omega r^2-\Omega_c r_c^2 \right) = 2 \pi \alpha P H r^2 - \int_{r_c}^r \frac{\mu^2}{r^4} \left 
         (1- \frac{\Omega}{\omega_s} \right) dr  ~~~~ ,
\eea
where we have taken the quantity $PH$ to vanish at $r_c$ under the assumption that the magnetic field quickly sweeps up any matter deposited inside $r_c$, thus maintaining a low pressure $P$.
For the $z$ component of the magnetic field we take $\mu/r^3$ to represent a dipole field whose axis is aligned approximately along the spin axis of the neutron star and perpendicular to the accretion disk. We further assume that $\Omega$ for all $r > r_c$ is always equal to the local Keplerian angular frequency.  

Before attempting to use equation (13) to solve for the disk structure, let us reflect a bit on its physical interpretation.  We start by rewriting equation (13) schematically as follows:
\bea
\dot J_{\dot M}(r) - \dot J_{\dot M}(r_c) = \Gamma_\alpha(r) - \Gamma_{B}(r) ~~~,
\eea
where the four terms are as follows: $\dot J_{\dot M}(r)$ is the angular momentum flux carried by matter across a boundary at radius $r$;  $\dot J_{\dot M}(r_c)$ is the angular momentum flux carried by matter through the inner edge of the disk and onto the neutron star; $-\Gamma_\alpha(r)$ is the net viscous torque on the disk interior to radius $r$; and $\Gamma_{B}(r)$ is the net magnetic torque on the disk interior to $r$.  If we move the $\Gamma_{B}(r)$ term to the other side of the equation and identify it as the angular momentum flux into the disk from the magnetic field of the neutron star, $\dot J_B$, we find:
\bea
\Gamma_\alpha(r) = \dot J_{\dot M}(r) - \dot J_{\dot M}(r_c) + \dot J_{B}(r) ~~~.
\eea
Written in this form, we can see that the viscous torques must transport angular momentum to the outer parts of the accretion disk at a rate equal to that transported by matter through $r$, minus the rate transported by matter through $r_c$, plus the rate flowing out of the neutron star and into the disk (inside $r$) via the magnetic field.  As $r$ approaches $r_c$ the magnetic torque vanishes and $\dot J_{\dot M}(r) \rightarrow \dot J_{\dot M}(r_c)$, thereby requiring zero viscous torques at the inner boundary ($P \rightarrow 0$ as well).  At the outer edge of the disk, presumably tidal torques from the companion star remove the angular momentum flow driven by $\Gamma_\alpha$, and feed it back into the orbit.  The net angular momentum flow into the neutron star is $\dot J_{\dot M}(r_c) - \dot J_{B}(r)$.  When this quantity becomes negative, the neutron star will be spun down.  The larger the magnetic field, the larger must be the viscous torques, $2\pi \alpha P H r^2$, in order to extract the increased angular momentum being injected into the disk.   This is accomplished by the disk readjusting its structure, i.e., the radial profiles of $P$ and $H$ must change in such a way as to transport the additional angular momentum toward the outer parts of the disk.  We show below that the principal enhancer of the viscous torque is a substantial increase in the pressure, i.e., the material becomes ``dammed up'' in the disk behind the corotation radius.

If we now carry out the integral representing the magnetic torque on the disk (eq. 13), and solve for the quantity $PH$, we find:
\bea
PH = \frac{\dot M \Omega}{2 \pi \alpha} \times F(r;r_c,\mu,\dot M)  ~~~~ ,
\eea
where $F$ is a dimensionless function of its argument and incorporates the effects of the magnetic torque on the disk.  Specifically, $F$ is:
\bea
F(r;r_c,\mu,\dot M) = \left(1-\sqrt{\frac{r_c}{r}}\right) + \frac{\xi^{7/2}}{9} \sqrt{\frac{r_c}{r}} 
\left\{1-3\left(\frac{r_c}{r}\right)^3+2\left(\frac{r_c}{r}\right)^{9/2}\right\} ~~~~ ,
\eea
where the first term in parentheses yields the SS solution in the absence of a magnetic field.  The parameter $\xi$ is just the ratio of $r_m/r_c$.

\begin{figure}[t]
\begin{center}
\includegraphics[height=5.0in,angle=-90]{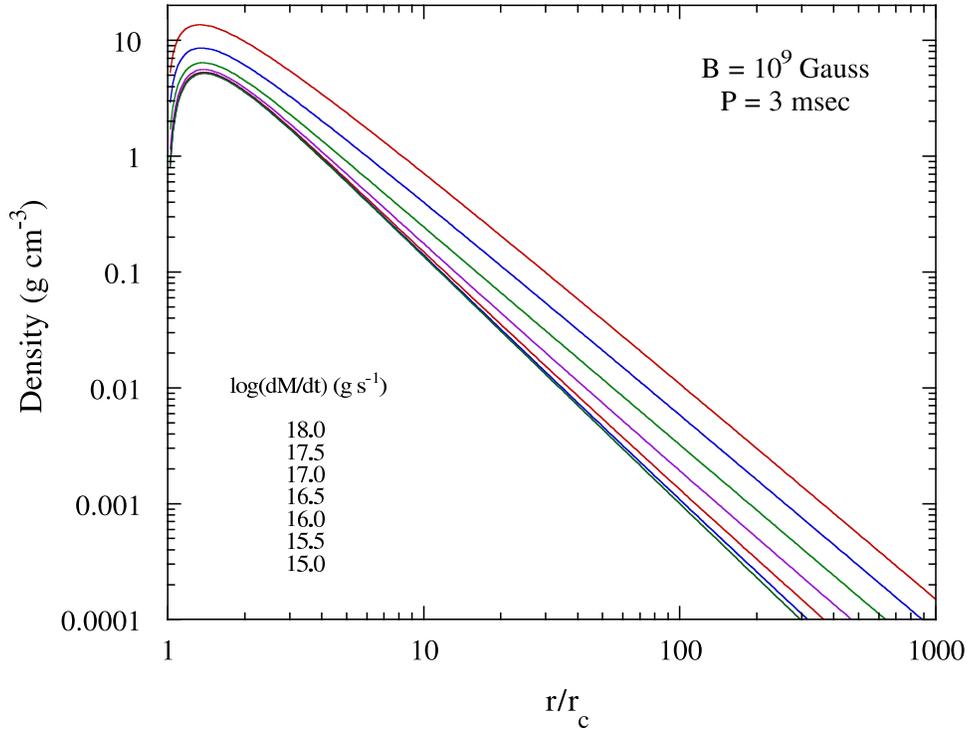}
\caption{The density $\rho$ as a function of radial distance, expressed in units of $r_c$. 
Curves are for different values of the assumed accretion rate $\dot M$, as indicated
by the set of labels; the highest curve is for the highest value of $\dot M$.  \label{fig:den}}
\end{center}
\end{figure}

Three of the four remaining SS-disk equations remain unchanged.  For the equation of state we take: $P=\rho k T/m$, where $m$ is the mean molecular weight per particle.  The vertical force balance equation is given by: $P \simeq GM\rho H^2 r^{-3}$.  Radiative transport in the vertical direction is represented by: $T^4 \simeq \kappa \rho H T_e^4$, where $T$ and $T_e$ are the midplane and effective temperatures of the disk, respectively, and $\kappa$ is the radiative opacity evaluated at the disk midplane.  Finally, we retain the fifth SS-disk equation for the heat dissipation per unit surface area of the disk: $\dot Q \simeq \alpha H P \Omega = \sigma T_e^4$.  In doing so, we have explicitly neglected the heat dissipation associated with the magnetic torque.  Since, in the disk solutions we are proposing, the viscous torques must overcome the magnetic torques (which tend to repel the disk material), the concomitant viscous heating term should dominate that due to magnetic heating.  For this reason, and in order to keep the disk equations in strictly algebraically solvable form, we neglect the magnetic heating.  In all of these equations (except for the equation of state) we have neglected dimensionless coefficients of order unity.  These factors have no impact on the form of the solutions for $P$, $\rho$, $H$, and $T$, and only a minor effect on the leading coefficients to the solutions.

The four equations listed above, plus the equation for $PH$ given in equation (16) can be solved algebraically in a manner analogous to that done for the original SS-disk equations to yield $P$, $\rho$, $H$, and $T$ as functions of $r$, $\dot M$, $\alpha$, and $\mu$.  In solving these equations we have taken $\kappa$ to be given by Kramers opacity ($\kappa \simeq 6 \times 10^{22} \rho T^{-3.5}$ cm$^2$ gm$^{-1}$) which is appropriate for most of the physical conditions found in our disk models, and the mass of the neutron star is fixed at $1.4~M_\odot$.  The solutions are:

\bea
P \simeq 2 \times 10^5 \alpha^{-9/10} \dot M_{\rm 16}^{17/20} r_{\rm 10}^{-21/8} F^{17/20}
~~~~~{\rm dynes~cm}^{-2}
\eea
\bea
H \simeq 1 \times 10^8 \alpha^{-1/10} \dot M_{\rm 16}^{3/20} r_{\rm 10}^{9/8} F^{3/20}
~~~~~{\rm cm}
\eea
\bea
T \simeq 2 \times 10^4 \alpha^{-1/5} \dot M_{\rm 16}^{3/10} r_{\rm 10}^{-3/4} F^{3/10}
~~~~~{\rm K}
\eea
\bea
\rho \simeq 7 \times 10^{-8} \alpha^{-7/10} \dot M_{\rm 16}^{11/20} r_{\rm 10}^{-15/8} F^{11/20}
~~~~~{\rm g~cm}^{-3}
\eea
where $\dot M_{\rm 16}$ is the mass accretion rate in units of $10^{16}$ gm sec$^{-1}$, and $r_{\rm 10}$ is the radial distance in units of $10^{10}$ cm.  Note that for a vanishing magnetic field, the function $F$ reduces to the function $f^4=1-\sqrt{r_c/r}$,  found in the SS-disk solutions (see eq. [17]).

As an example of how the magnetic torque on the disk affects the disk solutions, we show in Figure \ref{fig:den} plots of $\rho(r)$ for several different values of $\dot M$.  The plots are for the illustrative parameter values of a surface magnetic field  $B = 1 \times 10^9$ G, viscosity parameter $\alpha = 0.01$, and a spin period $P = 0.003$ sec.  The highest curve is for the value of $\dot M$ that corresponds approximately to the transition from a ``slow'' to ``fast'' pulsar; the lower curves are for progressively lower values of $\dot M$ and correspondingly ``faster pulsars''.   Note that for $r/r_c \gg 1$ the curves approach the standard SS solution with $\rho \propto \dot M^{11/20} r^{-15/8}$ (see eq.[21]).  However, nearer the corotation radius, the density for the lower $\dot M$ values rises more steeply and is systematically larger than it would be in the absence of the magnetic field. 

\begin{figure}[t]
\begin{minipage}[c]{0.47\textwidth}
\begin{center}
\includegraphics[height=0.97\textwidth,angle=-90]{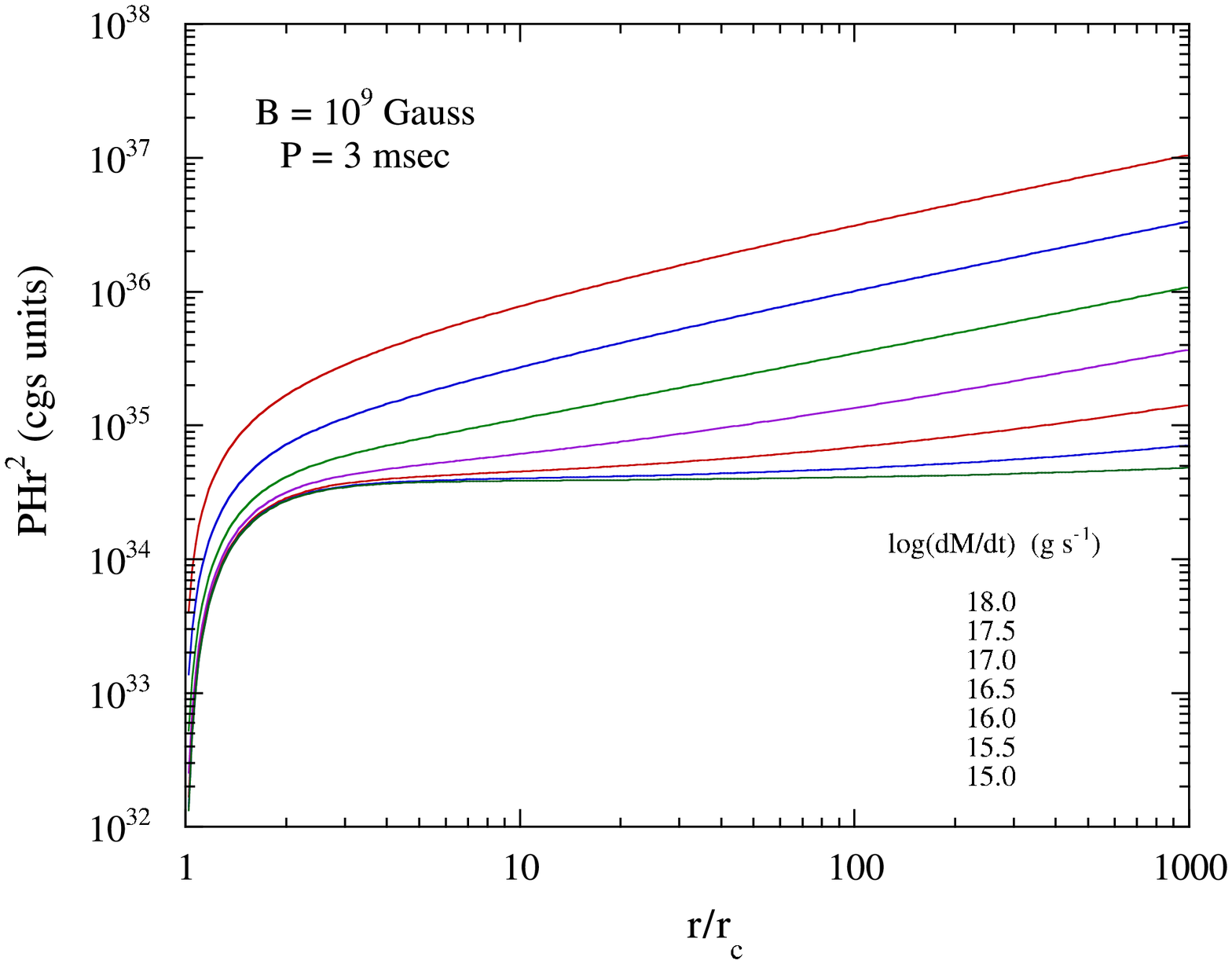}
\caption{The product $PHr^2$ as a function of radial distance, expressed in units of $r_c$. 
Curves are for different values of the assumed accretion rate $\dot M$, as indicated
by the labels. The curve for the highest (lowest) value of $\dot M$ is at the top (bottom). \label{fig:PH}}
\end{center}
\end{minipage}
\hfill
\begin{minipage}[c]{0.47\textwidth}
\begin{center}
\includegraphics[height=0.97\textwidth,angle=-90]{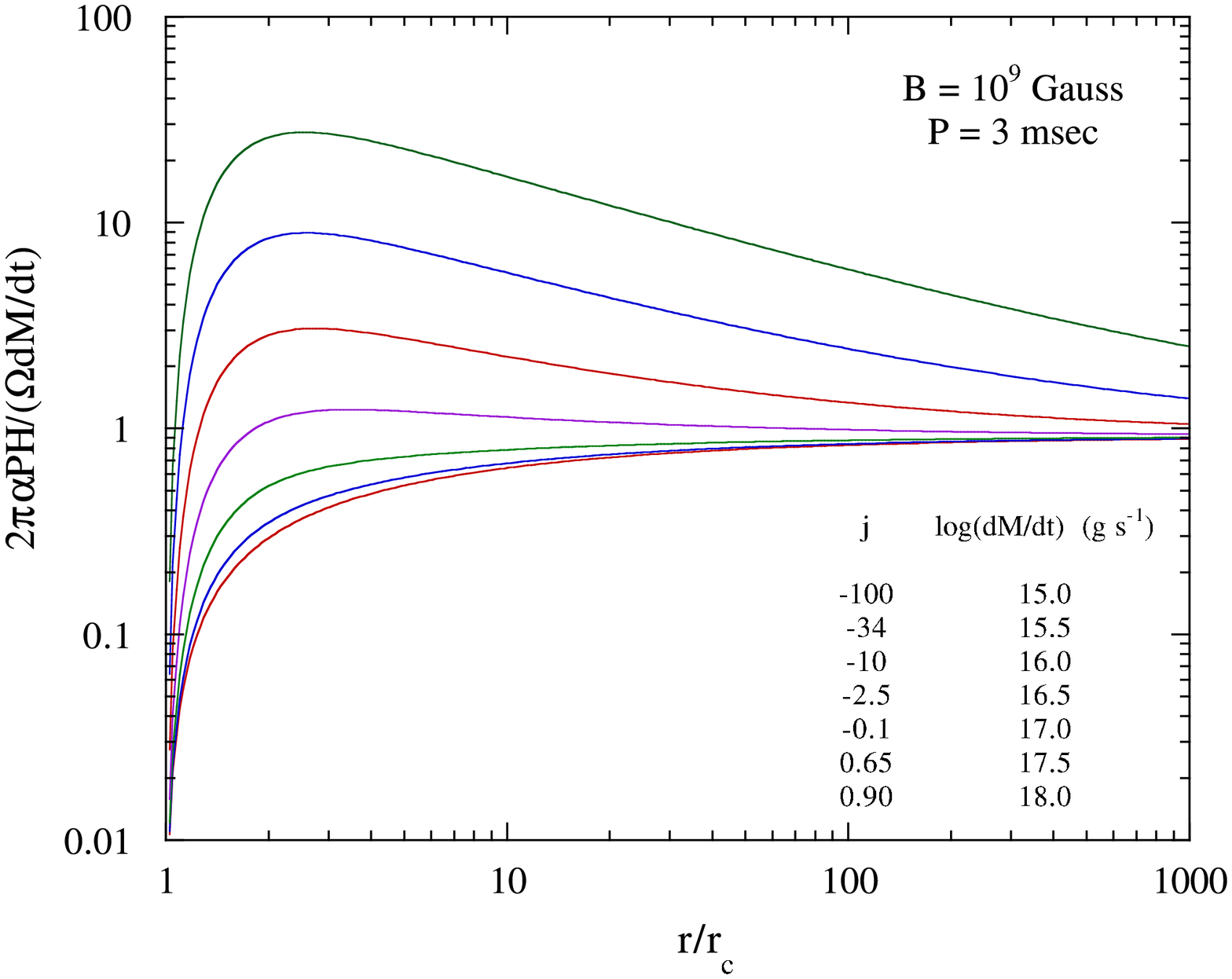}
\caption{The quantity $2 \pi \alpha PH/\Omega \dot M$ as a function of radial distance,
expressed in units of $r_c$.  This dimensionless ratio would equal $f^4=1-\sqrt{r_c/r}$ and be independent of $\dot M$ for the simple case of an SS disk. Note that in this figure, the curve 
for the lowest (highest) value of $\dot M$ is at the top (bottom), i.e., the reverse of the order
in Fig. \ref{fig:PH}.  Values of the relative angular momentum flux into the neutron star $j \equiv \dot J/(\dot M \Omega r^2)_{\rm c}$ are also indicated on the lower right. Negative values correspond to outward flux (spindown of the star) and high surface densities at the inner disk. 
\label{fig:PHOmeg}}
\end{center}
\end{minipage}
\end{figure}  

Since the one disk equation that we have modified (from the original SS-formulation) involves the term $PHr^2$ (see eq. [12]), we can gain a better understanding of the solutions by plotting this quantity (see Figure \ref{fig:PH}).  Note that the behavior of the $PHr^2$ curves shows a flattening for lower values of $\dot M$.   For low values of $\dot M$ what happens is that the viscous torques on the disk (1st term on the r.h.s. of eq. [13]) become relatively large and go almost exclusively to counteracting the magnetic torques (tending to drive the disk material outward; 2nd term on the r.h.s. of eq. [13]).  The difference between these two terms is quite small and provides a net drag on the disk material that drives the mass flow inward at a rate of $\dot M$ (l.h.s. of eq. [13]).  In this limit (of low $\dot M$), the $\xi^{7/2}$ term in eq. (17) dominates over the $(1-\sqrt{r_c/r})$ term, and $PHr^2 \propto \{1-3(r_c/r)^3+2 (r_c/r)^{9/2} \}$.  For $r$ only modestly larger than $r_c$, $PHr^2 \rightarrow$ constant, instead of growing approximately as $\sqrt{r}$ as in the slower pulsar case.  These effects are clearly seen in the plot of $PHr^2$  vs. $r$ shown in Fig. \ref{fig:PH}.

As a further comparison of these magnetically torqued disk solutions against the SS solutions, we plot in Fig. \ref{fig:PHOmeg} the quantity $2 \pi \alpha PH/(\Omega \dot M)~(\simeq 3\pi\nu\Sigma/\dot M$) which is a simple function of $r$ for an SS disk ($f^4=1-\sqrt{r_c/r}$), approaches unity as $r \gg r_c$, and is independent of $\dot M$.  Note that the curves which deviate most from unity (for intermediate and large $r$) are the solutions corresponding to the lower values of $\dot M$, i.e., the faster pulsars.

\section{Torques on the Neutron Star}

If, for fast X-ray pulsars, there exist accretion disks terminating at the corotation radius, as described above, then the net torque on the neutron star is straightforward to compute:
\bea
\Gamma_{\rm NS}\simeq \dot M \sqrt{GMr_c} - \int_{r_c}^\infty B_z^2(r) r^2 \left(1-\frac{\Omega}{\omega_s} \right) dr ~~~ 
\eea
(see also, LPP, GL, Wang 1987; 1995; 1996).  Formally, the integral should be carried out from $r_c$ to an upper limit equal to the radius of the
speed-of-light cylinder, $r_{\rm cyl}$.  However, for the msec pulsars of interest $r_{\rm cyl} \gg r_c$.  Equation (22) can be integrated to yield the torque on a fast X-ray pulsar of:
\bea
\Gamma_{\rm NS} \simeq \dot M \sqrt{GMr_c} - \frac{\mu^2}{9r_c^3} ~~~ .
\eea
 Note that the first term on the right represents the usual spin-up torque due to matter accreting from $r_c$, while the second term is a spin-down torque due to the magnetic field drag on the accretion disk.  The factor of 1/9 in the 2nd term on the r.h.s. of eq. (23) results from our particular choice for $B_\phi$ (see, eq. [11]); for other plausible functional forms this factor would change somewhat.  With our choice for $B_\phi$, the torque reverses sign when $r_m = 9^{2/7} r_c \simeq 1.87 r_c$; thus spinup can occur even while the pulsar is still `fast', i.e., $r_m \gtrsim r_c$.  In this regard, we remind the reader that the expression for $r_m$ given by eq.~(1) is a formal definition of the magnetospheric radius and is not calculated self-consistently.
 
In order to examine the behavior of the overall torque, $\Gamma_{\rm NS}$, as a function of $\dot M$, we need a prescription for the case where the pulsar switches from ``fast'' to ``slow'', i.e., where $r_m$ becomes smaller than $r_c$.  When this occurs we simply adopt the value of $r_m$ given by eq. (1) as the approximate location of the inner edge of the disk, and integrate the magnetic torque from $r_m$ to $\infty$.  By analogy with the ``ad hoc'' choice for $B_\phi$ that we made for $r > r_c$, we take $B_\phi = B_z (1-\omega_s/\Omega)$ for $r_m < r < r_c$.  Thus, for a slow pulsar, we find the torque on the neutron star to be:
\bea
\Gamma_{\rm NS} \simeq \dot M \sqrt{GMr_c} + \frac{\mu^2}{3r_c^3} \left[ \frac{2}{3}-2\left(\frac{r_c}{r_m}\right)^{3/2}+\left(\frac{r_c}{r_m}\right)^3 \right] ~~~ .
\eea

\begin{figure}[t]
\begin{center}
\includegraphics[height=5.0in,angle=-90]{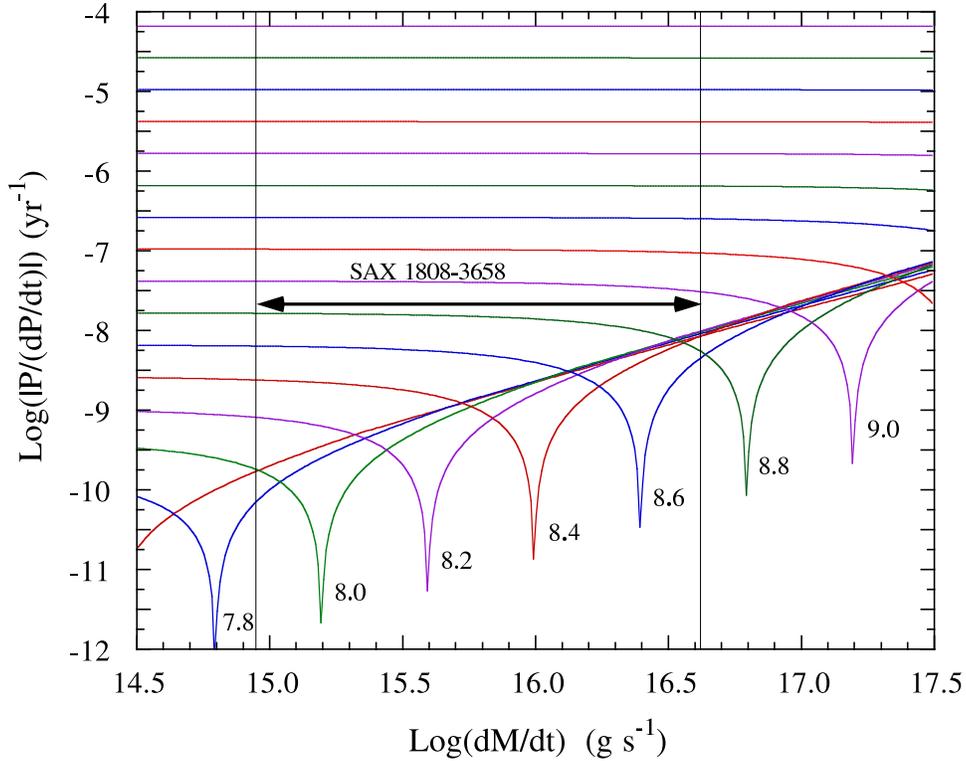}
\caption{Curves of $|\dot P/P|$ as a function of the accretion rate $\dot M$ for logarithmically spaced values of the assumed surface magnetic field of the neutron star (in units of G). The pulse period is taken to be 2.5 msec to represent the case of SAX 1808-3568.  \label{fig:torque}}
\end{center}
\end{figure} 
 
 In Figure \ref{fig:torque} we show the results of evaluating $|\dot P/P|$ as a function of $\dot M$ for a range of surface magnetic fields. In order to convert the torque, given in eqs. (23) and (24), to $\dot P/P$ one needs the rotation frequency of the neutron star.  For this we have chosen an illustrative spin period of 2.5 msec, i.e., the value for the SAX 1808-3658 pulsar.  The cusp-like structure in each curve is the location where $\dot P$ changes sign.  The vertical bars show the approximate range of $\dot M$ inferred for SAX 1808-3658 during its most recent outburst in 2002 (for a distance of $\sim 4$ kpc; Chakrabarty \& Morgan 1998; Homer et al. 2001).  The heavy arrow is the average value of $\dot P/P \simeq 2 \times 10^{-8}$ yr$^{-1}$ observed over a period of $\sim$4 weeks during the 2002-outburst while the source declined in intensity by a factor of $\sim$50 (Morgan, Galloway, \& Chakrabarty 2003).   By inspection of Figure \ref{fig:torque} it can be seen that the proposed model indicates a surface $B$ field of $\sim 8 \times 10^8$ G.  We point out that the magnitude of the observed torque is substantially larger than that expected for the matter torque alone over most of the duration of the outburst cycle (especially for the lower values of $\dot M$).   In this respect, the proposed model can successfully explain the observed spin period behavior -- especially if the magnetic field is ultimately confirmed to be as large as $\sim 8 \times 10^8$ G.  We also note, however, that a reanalysis of the spin behavior of SAX 1808-3658 during its 1998 outburst (Morgan, Galloway, \& Chakrabarty 2003) has detected a significant {\em spinup} over most of its on state---and of a magnitude that is even larger than that of the recently observed spindown.   Thus, we caution that the msec pulsars are apparently capable of exhibiting spinups and spindowns that are inconsistent with the inferred value of $\dot M$, as is the case for their more slowly rotating cousins (see, e.g., Nelson et al. 1997).
 
XTE J0929-314 has thus far been observed only during a single outburst (Galloway et al. 2002).  During that time the source underwent spindown with a value of $\dot P/P \simeq 1.6 \times 10^{-8}$ yr$^{-1}$, and had a luminosity $L_x < 1.5 \times 10^{36}$ ergs s$^{-1}$ (for an assumed distance of 5 kpc; Galloway et al. 2002).  An analysis similar to that done for SAX 1808-3658 (Fig. \ref{fig:torque}) yields a comparable value for the magnetic field of $\sim 1.0 \times 10^9$ G.  Accretion torques have not yet been reported for the other three known msec X-ray pulsars.  

\newpage
   
\section{Discussion and Conclusions}

In this work, we have explored the proposition that fast X-ray pulsars can continue to accrete from a thin disk, even for accretion rates that place the nominal magnetospheric radius well beyond the corotation radius.  We hypothesize that the inner edge of the accretion disk will be located just inside the corotation radius.  We modify the Shakura-Sunyaev disk equations by adding a simple magnetic torque prescription, and find analytic solutions to these equations.  The density and pressure profiles are significantly modified over the SS solutions as the fastness parameter increases.  We have not analyzed the physical stability of these solutions, and therefore cannot comment on the lower limit on $\dot M$ which would be allowed in our model.  The form of the magnetic torques we have assumed are admittedly subject to substantial uncertainty; however, the main conclusions we reach, e.g., accretion over a wide range of $\dot M$ and compatibility of {\em spindown} during accretion should be largely independent of the exact form adopted (as argued before in ST).

The proposed model directly confronts the observational fact that the accretion by millisecond X-ray pulsars apparently persists over a wide range of $\dot M$, including down to very low values of $\dot M \lesssim 10^{-11} M_\odot$ yr$^{-1}$.  The model can also account for the magnitude of the observed spindown torques that have been observed for two of the msec X-ray pulsars.  It does not easily explain the magnitude of the {\em spinup} episode in SAX J1808-3658 during its 1998 outburst.   The observed spindown behavior in SAX J1808-3658 (during the 2002 outburst) and XTE J0929-314 requires surface magnetic fields of $\sim 8 \times 10^8$ G; this is a prediction that can possibly be used to verify or falsify the model.  Also, if correct, our model would remove the constraints on the neutron star radius in SAX J1808-3658 found by Burderi \& King (1998) and Psaltis \& Chakrabarty (1999). 

Campana et al. (2001) reported a rapid transition in the X-ray luminosity of 4U 0115+63 by a factor of $\gtrsim$~250 over a 15-hr interval, during which time the pulsations ($P=3.6$ sec) continued with only minor changes in pulse fraction.  This change in luminosity, according to the standard accretion picture, should have signaled the onset of a transition from direct accretion onto the neutron star surface to a propeller mode.  Campana et al. (2001) developed a simple model to account for this behavior without invoking the type of ``fast-pulsar accretion'' advocated in this work.  In their model, the actual value of $\dot M$ approaching $r_m$ changes by only a factor of a few, while the luminosity changes by a factor of $\gtrsim$~250.  Based on the results presented, it is difficult to judge whether this model is to be preferred over the one presented herein.  First, we note that these authors did not allow for the possibility of direct, continuous accretion onto a fast pulsar.  Second, they assert that the rapid change in luminosity over a 15-hr interval cannot naturally be accounted for by variations in $\dot M$ flowing through the disk.  However, given our lack of detailed understanding of Be star excretion disks, it does not seem completely implausible that the change in X-ray luminosity directly reflects the change in $\dot M$ into the disk.  Finally, interesting predictions of both models would be the sign and magnitude of the spin-torque behavior of the neutron star---however, no such measurements were reported.

Twin-peak kHz quasiperiodic oscillations (``QPOs'') have been observed in the X-ray intensity of two of the msec pulsars (SAX 1808-3658 and XTE J1807-294; Chakrabarty et al. 2003; Markwardt et al. 2003).  In some models, one of the two kHz QPO peaks represents the orbital frequency of blobs of matter near the neutron star (see, e.g., Miller, Lamb, \& Psaltis 1998; Stella \& Vietri 1999).  We note that such high Keplerian frequencies cannot be reached outside of $r_c$ where the accretion disk is located in our model.  If such an association between the kHz QPOs and Keplerian motion is firmly established, our model would not naturally accommodate these orbital frequencies; however, we emphasize that the kHz QPOs are not well understood at this time.

At an accretion rate of $\dot M = 2 \times 10^{15}$ gm s$^{-1} \simeq 3 \times 10^{-11} M_\odot$ yr$^{-1}$ our disk model requires $\sim 2$ yr to fill from an empty state; the corresponding filling time of an SS disk is $\sim 1$ yr.  The difference of $\sim 1$ yr is comparable to the recurrence time of these transient msec pulsars, and may in some way be related to the transient outbursts.  This aspect of the model should be pursued with time-dependent calculations.  

\acknowledgements
We thank Alan Levine and Eric Pfahl for helpful discussions.  Edward Morgan and Deepto Chakrabarty supplied important results on the spin behavior of SAX 1808-3658 prior to their publication.  James Kiger and John Belcher participated in an earlier version of this research.  One of us (SR) acknowledges support from NASA ATP Grant NAG5-12522.

\end{document}